\documentclass[twocolumn,showpacs,aps,prd,amsmath,amssymb,amsfonts,%
superscriptaddress,nofootinbib]{revtex4-1}
\usepackage{graphicx}
\usepackage{dcolumn}
\usepackage{bm}

\begin{document}
\newcommand*{\ea}{\textit{et al.}}
\newcommand*{\zpc}[3]{Z.~Phys.~C \textbf{#1}, #2 (#3)}
\newcommand*{\plb}[3]{Phys.~Lett.~B \textbf{#1}, #2 (#3)}
\newcommand*{\pr}[3]{Phys.~Rep.~\textbf{#1}, #2 (#3)}      
\newcommand*{\npa}[3]{Nucl.~Phys.~A \textbf{#1}, #2 (#3)}  
\newcommand*{\npb}[3]{Nucl.~Phys.~B \textbf{#1}, #2 (#3)}  
\newcommand*{\npbps}[3]{Nucl.~Phys.~B (Proc. Suppl.) \textbf{#1}, #2 (#3)}  
\newcommand*{\rmph}[3]{\rmp~\textbf{#1}, #2 (#3)}
\newcommand*{\ibid}[3]{\textit{ ibid.} \textbf{#1}, #2 (#3)}
\newcommand*{\sjnp}[4]{Yad. Fiz. \textbf{#1}, #2 (#3) [Sov. J. Nucl. Phys.
\textbf{#1}, #4 (#3)]}
\newcommand*{\epjc}[3]{Eur. Phys. J. C \textbf{#1}, #2 (#3)}
\newcommand*{\tauch}{\tau^-\ra\nu_\tau\pi^-\pi^+\pi^-}
\newcommand*{\taunn}{\tau^-\ra\nu_\tau\pi^-\pi^0\pi^0}
\newcommand*{\tripi}{\pi^-\pi^+\pi^-}
\newcommand*{\pitwo}{\pi^-\pi^0\pi^0}
\newcommand*{\nl}{\nonumber \\}
\newcommand*{\bea}{\begin{eqnarray}}
\newcommand*{\eea}{\end{eqnarray}}
\newcommand*{\barr}{\begin{array}}
\newcommand*{\earr}{\end{array}}
\newcommand*{\bi}{\bibitem}
\newcommand*{\be}{\begin{equation}}
\newcommand*{\ee}{\end{equation}}
\newcommand*{\rmrs}{M_\rho^2(s)}
\newcommand*{\mrs}{m_\rho^2}
\newcommand*{\ra}{\rightarrow}
\newcommand*{\die}{e^+e^-}
\newcommand*{\arp}{{a_1\rho\pi}}
\newcommand*{\akstk}{{a_1K^*K}}
\newcommand*{\eg}{e.g.}
\newcommand*{\amu}{{\mathbf A}^\mu}
\newcommand*{\anu}{{\mathbf A}^\nu}
\newcommand*{\amunu}{{\mathbf A}^{\mu\nu}}
\newcommand*{\vmu}{{\mathbf V}_\mu}
\newcommand*{\vmunu}{{\mathbf V}_{\mu\nu}}
\newcommand*{\fai}{\mathbf P}
\newcommand*{\rf}[1]{(\ref{#1})}
\newcommand*{\rmas}{M^2_{a_1}(s)}
\newcommand*{\mas}{m_{a_1}^2}
\newcommand*{\lag}{{\mathcal L}}
\newcommand*{\sth}{\sin\theta}
\newcommand{\opava}{Institute of Physics, Silesian University in Opava,
Bezru\v{c}ovo n\'{a}m. 13, 746 01 Opava, Czech Republic}
\newcommand{\praha}{Institute of Experimental and Applied Physics,
Czech Technical University, Horsk\'{a} 3/a, 120 00 Prague, Czech
Republic}

\title{Three-pion decays of the tau lepton, the $\mathbf{a_1(1260)}$ 
properties, and the $\bm{a_1\rho\pi}$ Lagrangian}

\author{Martin Voj\'{\i}k}
\affiliation{\opava}
\author{Peter Lichard}
\affiliation{\opava}
\affiliation{\praha}

\date{\today}

\begin{abstract}
We show that the $a_1\rho\pi$ Lagrangian is a decisive element for 
obtaining a good phenomenological description of the three-pion decays 
of the $\tau$ lepton. We choose it in a two-component form with a 
flexible mixing parameter $\sin\theta$. In addition to the dominant 
$a_1\rightarrow\pi\rho$ intermediate states, the 
$a_1\rightarrow\pi\sigma$ ones are 
included. When fitting the three-pion mass spectra, three data sets 
are explored: (1) ALEPH 2005 $\pi^-\pi^+\pi^-$ data, (2) ALEPH 2005 
$\pi^-\pi^0\pi^0$ data, and (3) previous two sets combined and
supplemented with the ARGUS 1993, OPAL 1997, and CLEO 2000 data. 
The corresponding confidence levels are (1) 28.3\%, (2) 100\%, and (3)
7.7\%. After the inclusion of the $a_1(1640)$ resonance, the agreement 
of the model with data greatly improves and the confidence level 
reaches 100\% for each of the three data sets.  From the fit to all 
five experiments [data set (3)], the following parameters of the 
$a_1(1260)$ are obtained: $m_{a_1}=(1233\pm18)$~MeV, 
$\Gamma_{a_1}=(431\pm20)$~MeV. The optimal value of the Lagrangian 
mixing parameter $\sin\theta=0.459\pm0.004$ 
agrees with the value obtained recently from the $e^+e^-$ annihilation 
into four pions. 
\end{abstract}

\maketitle
\section{\label{introduc}Introduction}
The dominance of the $a_1(1260)$ meson, hereafter referred as $a_1$,
and its $\pi\rho$ decay mode in the 
three-pion decays of the $\tau$-lepton is firmly established experimentally 
\cite{pluto,delco1986,markii1986,argus1986,mac,cello,argus1993,opal1995,%
opal1997,aleph1998,delphi,cleo2000a,cleo2000b,cleo_prelim,aleph2005,argus1996}.
A convincing demonstration was provided by the ARGUS Collaboration 
\cite{argus1993,argus1996}, who 
compared the distribution of unlike- and like-sign two-pion masses. 
The $\pi\rho$ intermediate state is the core of several
models of the three-pion decays of the $\tau$-lepton 
\cite{pham,tornqvist1987,isgurtau,kuhnsanta,feindt,kuhnmirkes,poffenberger,%
dpp,drpp,achasov,cleo2000a}. 
It should be mentioned that the $a_1$ dominance in heavy 
lepton decays was proposed in 1971 \cite{tsai}, about five years before 
the $\tau$-lepton was actually discovered.

The  $a_1(1260)$ resonance, discovered almost fifty years ago
\cite{goldhaber}, plays an important role in many phenomena of 
the nuclear and particle physics. Its properties have been studied in 
many processes, but even its basic parameters are not very well known.
The values of the $a_1(1260)$ resonance mass determined from different 
processes or by different experimental groups often contradict one another. 
The same applies, even to a larger extent, to the $a_1$ width. 
Very little improvement has been achieved over the last thirty years, see 
Table \ref{tab:pdg}.
\begin{table}
\caption{\label{tab:pdg}Thirty-year history of the basic $a_1(1260)$ 
parameters as listed in the Particle Data Group Tables. Only the last 
edition and those in which a change appeared are shown.}
\begin{ruledtabular}
\begin{tabular}{cccc}
PDG  & $m_{a_1}$~(MeV) & $\Gamma_{a_1}$~(MeV) & $a_1\ra\rho\pi$\\
\colrule
1978 \cite{pdg1978}& $\approx$ 1100& $\approx$ 300 & $\approx$ 100\%\\
1980 \cite{pdg1980}& 1100 to 1300& $\approx$ 300 & dominant\\
1982 \cite{pdg1982}& 1275$\pm$30& 315$\pm$45 & dominant\\ 
1986 \cite{pdg1986}& 1275$\pm$28& 316$\pm$45 & dominant\\   
1988 \cite{pdg1988}& 1260$\pm$30& 300 to 600 & dominant\\   
1990 \cite{pdg1990}& 1260$\pm$30& 350 to 500 & dominant\\    
1992 \cite{pdg1992}& 1260$\pm$30& $\approx$ 400 & dominant\\ 
1994 \cite{pdg1994}& 1230$\pm$40& $\approx$ 400 & dominant\\ 
1998 \cite{pdg1998}& 1230$\pm$40& 250 to 600 & dominant\\
2000 \cite{pdg2000}& 1230$\pm$40& 250 to 600 & seen\\
2008 \cite{pdg2008}& 1230$\pm$40& 250 to 600 & seen\\
\end{tabular}
\end{ruledtabular}
\end{table}
The origin of those problems lies in the very nature of the $a_1$ resonance
with its short lifetime and large width. The usual definition of the
mass and usual procedures for its measurement are not applicable. The $a_1$
mass and width enter the formulas for experimentally accessible quantities 
via the assumed form of the resonance propagator, which generates a specific
Breit-Wigner formula. Those formulas are further modified in different ways
when modeling the dynamics of the processes in which the $a_1$ participates. 
As a result, we do not have a unique definition of the $a_1$ mass $m_{a_1}$ 
and width $\Gamma_{a_1}$. In fact, every formula represents a specific 
definition of $m_{a_1}$ and $\Gamma_{a_1}$. Given this, it is not surprising 
that different models yielded different results even when being applied to 
the same data. It would be natural to accept as the $a_1$ canonical 
parameters the results of a model that best describes a broad class of data 
on various processes and from various experiments. Unfortunately, we are not 
in such a situation yet.

The situation of the heavier meson states with $J^{PC}=1^{++}$ is 
a little unclear and none of them has found its place in the Summary Table
of the Review of Particle Properties \cite{pdg2008}. The first indication 
of the state with mass of 1.65~GeV and width of 0.4~GeV appeared already 
in 1978 \cite{pernegr1978}. The later experimental evidence, which comes 
mainly from hadronic reactions, is summarized in \cite{pdg2008}, where 
this resonance is listed as $a_1(1640)$ and assigned the mass of 
$(1647\pm22)$~MeV and the width of $(254\pm27)$~MeV.
The three-pion decay of the $\tau$-lepton is less convenient for 
studying the $a_1(1640)$ resonance (often denoted as $a_1^\prime$ in what 
follows) because of fundamental limitations due to the $\tau$ mass 
which is not big enough to provide sufficient phase space for three-pion 
final states with the needed invariant mass. Nevertheless, the DELPHI
Collaboration \cite{delphi} performed the analysis of the Dalitz
plots for different 3-pion mass ranges and observed an enhancement that
``could be explained by a decay mode of the $\tau$ 
to a resonance of mass similar to or greater than the $\tau$ mass which 
then decays to three pions through the intermediate state of a pion plus a
particle of mass 1.25 GeV or greater.'' They interpreted this as an evidence
for the $a_1^\prime$. In our opinion, this observation need not signify 
the existence of the $a_1^\prime$. It may also be a decay of the $a_1(1260)$, 
produced with a larger than nominal mass, into $\pi$ and $\rho(1450)$. 
A more convincing proof of the $a_1^\prime$ in the decay of the $\tau$ 
lepton comes from the CLEO Collaboration \cite{cleo2000a}. They showed 
that adding the $a_1^\prime$ term into the Breit-Wigner function improved 
significantly the agreement with the data.

On the theoretical side, a radial excitation of the quark-antiquark system 
with a mass of 1.82~GeV appeared in a relativized quark model with 
chromodynamics of Godfrey and Isgur \cite{godfrey1985}. Its decay width into 
the $\pi\rho$ channel was calculated in the flux-tube-breaking model 
by Kokoski and Isgur \cite{kokoski1987} with result $\alt 70$~MeV (our 
estimate is based on their Table II). The seminal analysis of Barnes, 
Close, Page, and Swanson \cite{barnes1997} has shown that the experimentally 
observed dominance of the D-wave over S-wave \cite{ves1995,adams1998} 
excludes the hybrid meson nature of the $a_1^\prime$ and confirms it as 
a radial excitation of the quark-antiquark system. 

A few $I^G(J^{PC})=1^-(1^{++})$ meson states above the $a_1(1640)$ have been 
observed by a single group, mainly in the 
$p\bar{p}$ annihilation. They still need confirmation. For details, 
see \cite{pdg2008}.

Another important ingredient that defines a particular model of the 
three-pion decay of the $\tau$-lepton is, besides the $a_1$ propagator, 
the $\arp$ vertex.  The models assembled by different authors are
based on different $\arp$ vertexes. These are sometimes simply constructed 
as allowed combinations of the metric tensor and participating
four-momenta. A more rigorous way lies in deriving them from the interaction 
Lagrangians among the axial, vector and pseudoscalar fields. Unfortunately,
here the situation is unclear yet. Various theoretical concepts provide 
different effective Lagrangians \cite{theory}.
This is probably the reason why the model builders 
preferred trivial Lagrangians or \textit{ad hoc} vertexes. However, recent 
articles \cite{dpp,drpp,achasov} are different. Dumm, Pich, and
Portol\'{e}s \cite{dpp} got their Lagrangian from the resonance chiral
theory. Their work was revised in the light of later developments in
\cite{drpp}. Achasov and Kozhevnikov \cite{achasov} used the Generalized
Hidden Local Symmetry model.

Several models of the three-pion decay of the tau lepton have been proposed.
With some simplification one can say that each of them gives compatible 
results when applied to 
different sets of data, but the results of different models are incompatible.
Also the agreement of many models with data (often verbally claimed as
satisfactory) is poor when judged by usual statistical criteria.
The most popular models were those of Isgur, Morningstar, and Reader (IMR) 
\cite{isgurtau} and of K\"{u}hn and Santamaria (KS) \cite{kuhnsanta}. 
Other models were much less successful in fitting the data. As an example 
we recall the results from \cite{argus1993}, where the ARGUS Collaboration
compared various models with their data. Using the $\chi^2$'s and the
numbers of degrees of freedom (NDF) from their Table 4, we are
getting the confidence level (C.L.) of $\approx10^{-4}$ for Bowler's model 
\cite{bowler1986} and 2.2\% for the model of Ivanov, Osipov, and Volkov 
\cite{ivanov1991}. The KS and IMR models look better with C.L. 10.7\% and 
79.0\%, respectively. However, in a later article \cite{argus1995} the ARGUS
Collaboration used an enlarged set of data (integrated luminosity of
445~pb$^{-1}$ against 264~pb$^{-1}$ in \cite{argus1993}) and found that the 
KS model is rejected on a 7.4~$\sigma$ level. The IMR model with parameters
as given in \cite{isgurtau} was incompatible with the data on the same
level \cite{argus1995}.

Up to now, the best results have been obtained by the CLEO model 
\cite{cleo2000a} and by the model of Achasov and Kozhevnikov \cite{achasov}.
The former obtained, when fitting the CLEO $\pitwo$ data \cite{cleo2000a}, 
C.L. of 54.6\% without the $a_1^\prime$ resonance and 88.2\% with it. The 
latter fitted the ALEPH $\tripi$ data \cite{aleph2005} assuming two
heavier axial mesons $a_1^\prime$ and $a_1^{\prime\prime}$ and got 
$\chi^2$/NDF=79/102, which corresponds to C.L. of 95.6\%. Unfortunately, 
each of those two successful models has been applied only to one data set.

The finding of an $\arp$ Lagrangian that leads to a satisfactory description
of the three-pion production in the tau decays would have important
consequences for other areas of the high energy and nuclear physics.
For example, the $a_1$ resonance and its coupling to the $\rho\pi$ system
play important role in the evaluation of the dilepton and photon production
rates from a hadronic fireball presumably created in the relativistic heavy 
ion collisions. The calculations performed so far, see, \eg, Refs. 
\cite{elmag}, have shown that the yield of electromagnetic signals strongly 
depends on the choice of the $\arp$ Lagrangian. Fixing its correct form is 
thus important for distinguishing the electromagnetic radiation of the 
Quark-Gluon Plasma (QGP) from the hadronic sources. 

The outline for this paper is as follows. In Sec.~\ref{sec:model} we
describe our model and mention briefly its similarities and differences
with other models. The experimental data used for testing our model and 
fixing its parameters are listed in Sec. \ref{sec:data}.
Some details about our calculations and the results are presented in
Sec.~\ref{sec:results}. We summarize our results and conclude in
Sec.~\ref{sec:summary}. Two Appendixes contain technical details.
The present work supersedes an earlier paper \cite{licvoj}.

\section{\label{sec:model}Model of the three--pion decays of the tau lepton}
\begin{figure}
\includegraphics[width=8.6cm]{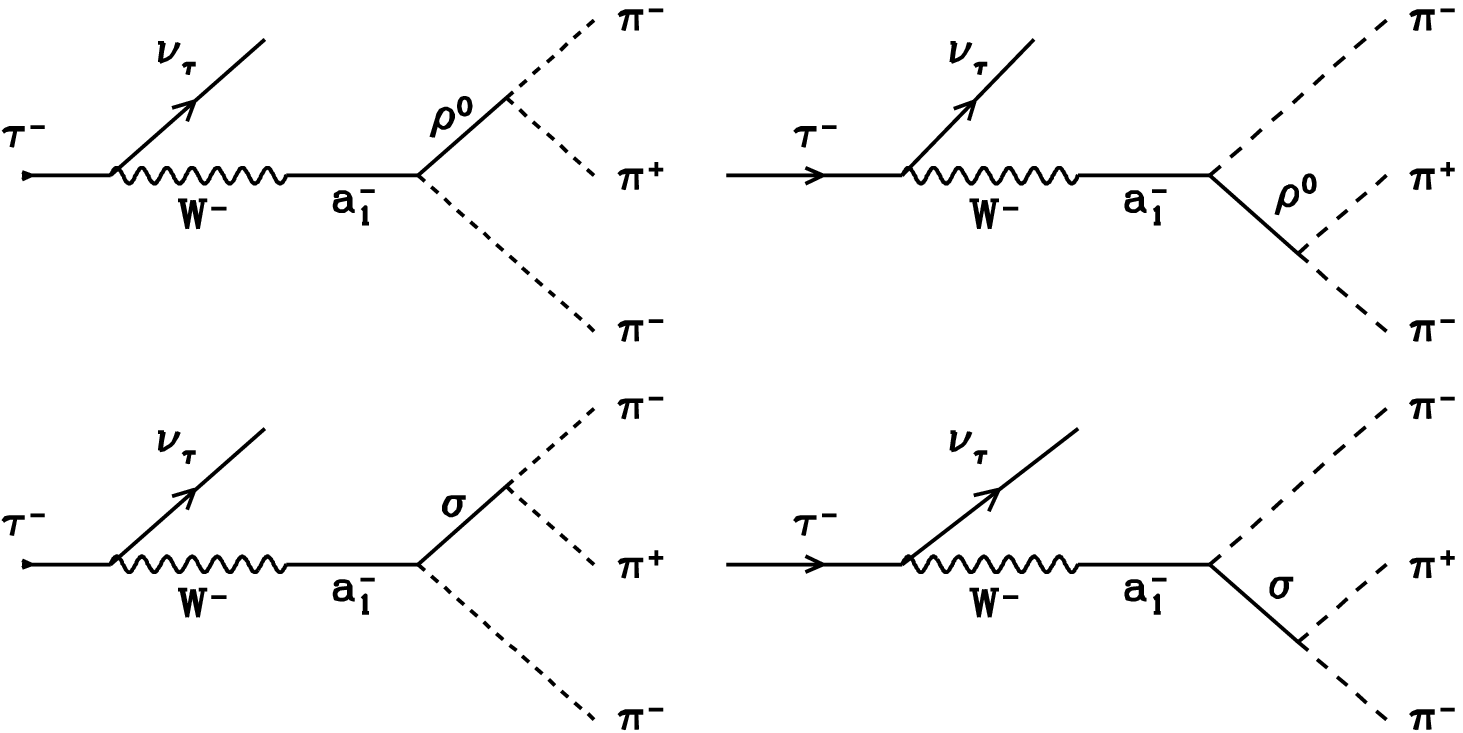}
\caption{\label{fig:fig1}Feynman diagrams of the $\tauch$ decay.}
\end{figure}

\begin{figure}
\includegraphics[width=8.6cm]{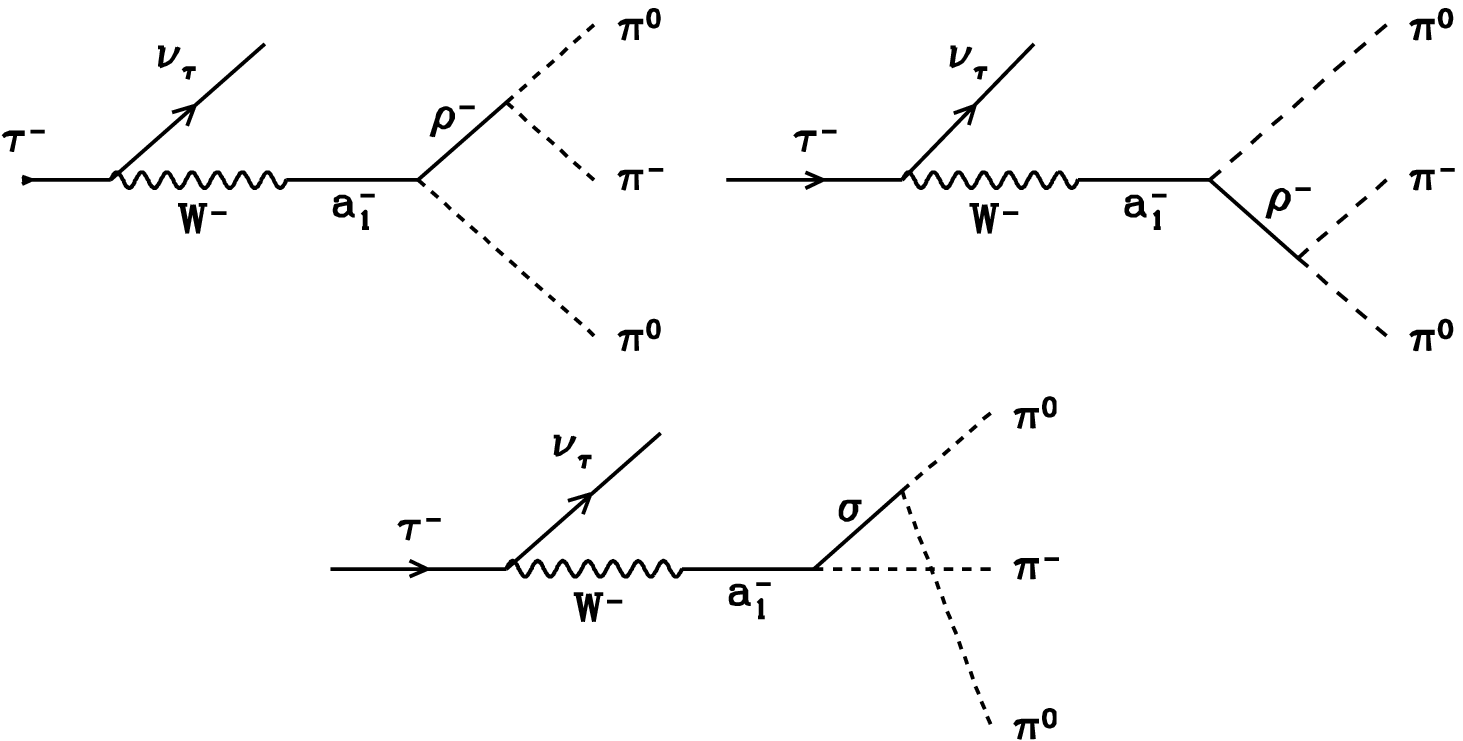}
\caption{\label{fig:fig2}Feynman diagrams of the $\taunn$ decay.}
\end{figure}

In this section we present our model, which will be used for fitting 
the three--pion mass spectra of the decays $\tauch$ and $\taunn$. The basic 
information can be obtained by inspecting Figs.~\ref{fig:fig1} and 
\ref{fig:fig2}. In addition to the standard $a_1\ra\pi\rho$ 
intermediate states we include also the states in which the $a_1$ couples 
to a pion and an $f_0(600)$ (hereafter called $\sigma$). The $\pi\sigma$ 
intermediate states improve the behavior of the differential
decay width at small masses of the three-pion system and bring the
difference between $\tauch$ and $\taunn$ decays. Their importance
has been pointed out by the CLEO Collaboration \cite{cleo2000a}.

To show what is specific for our model, what differs it from other 
existing models of the three-pion decays of the tau lepton, we have 
to provide more information. This is done in the following subsections.

\subsection{\label{sec:lag}Phenomenological $\bm{\arp}$ Lagrangian}
The interaction Lagrangian among the $a_1$, $\rho$, and pion fields
implies the form of the $\arp$ vertex in the Feynman diagrams. But sometimes
a vertex is postulated that can hardly be related to any effective 
Lagrangian. In the literature, one can find several prescriptions 
for the $\arp$ vertex used in the calculation of the decay rate of the tau 
lepton into three pions and neutrino. The simplest one is 
$X^{\alpha\mu}\propto g^{\alpha\mu}$, where index $\alpha$ ($\mu$) couples 
to the $a_1$ ($\rho$) line. It can be derived from the interaction 
Lagrangian among the $a_1$, $\rho$, and $\pi$ fields without derivatives. 
It was used, e.g., in Ref.~\cite{pham}. On the opposite pole of complexity 
is a two-component vertex used in the IMR model \cite{isgurtau}. Both its 
components are transversal both to the $a_1$ and $\rho$ four-momenta. The 
relative weight of the two components can vary, what gives the IMR model 
more flexibility. This is probably the main reason why this model sometimes 
fits the data a little better than the KS model \cite{kuhnsanta}, see,  
for example, \cite{opal1995}.

To maintain both the flexibility and the correspondence with the effective 
field theory, we use a two-component Lagrangian of the $\arp$ 
interaction in the form 
\be
\label{lagarp}
\lag_\arp=\frac{g_{\arp}}{\sqrt{2}}
\left(\lag_1\cos\theta+\lag_2\sin\theta\right),
\ee
where
\begin{eqnarray*}
\lag_1 &=& \amu\cdot\left(\vmunu\times\partial^\nu{\fai}\right),\\
\lag_2 &=& \vmunu\cdot\left(\partial^\mu\anu\times{\fai}\right),
\end{eqnarray*}
and $\vmunu = \partial_\mu{\mathbf V}_\nu-\partial_\nu\vmu$.
The isovectors $\amu$, $\vmu$, and $\fai$ denote the operators of 
the $a_1$, $\rho$ and $\pi$ fields, respectively. 

Our Lagrangian differs from that derived by Wess and Zumino, see 
Eq.~(67) in \cite{wz}, only by notation. We will consider the mixing 
angle $\theta$ a free parameter that has to be determined by fitting 
the experimental three-pion mass distribution. For each $\theta$, the 
coupling constant $g_\arp$ can be determined from the $a_1\ra\rho\pi$ 
decay width. The Lagrangian (\ref{lagarp}) implies the following 
$\arp$ vertex
\begin{eqnarray*}
X^{\alpha\mu} &=& \frac{ig_\arp}
{\sqrt{2}}\left\{cos\theta\left[p^{\alpha}_{\rho}p^{\mu}_{\pi}-\left(p_{\pi}
p_{\rho}\right)g^{\alpha\mu}\right]\right.\\
&-& \left. \sin\theta\left[p^{\alpha}_{\rho}
p^{\mu}_{a_1}-\left(p_{a_1}p_{\rho}\right)g^{\alpha\mu}\right]\right\},
\end{eqnarray*}
where $p$'s denote the four-momenta of the corresponding mesons (incoming
$a_1$, outgoing $\rho$ and $\pi$).

Lagrangian \rf{lagarp} has recently been used \cite{licjur,jurlic} in
a model of the electron--positron annihilation into four pions. Value
of the mixing parameter $\sth$ was obtained by fitting the excitation
function (dependence of the annihilation cross section on the invariant
collision energy). From the $\pi^+\pi^-\pi^+\pi^-$ channel the value
of $0.460\pm0.003$ has been obtained \cite{licjur}. In \cite{jurlic},
a combined fit to $\pi^+\pi^-\pi^+\pi^-$ and $\pi^+\pi^-\pi^0\pi^0$ 
channels has provided the value of $0.466\pm0.005$. 

\subsection{\label{sec:otherlags}Other effective Lagrangians and their 
parameters}
The Lagrangian describing the interaction of the $a_1$ triplet 
with strange pseudoscalar and vector mesons is not required when 
calculating the amplitudes of the three-pion decays of the $\tau$, 
see Figs. \ref{fig:fig1} and \ref{fig:fig2}. It is needed 
for evaluating the strange channel contribution to the total decay width 
of the $a_1(1260)$. The latter enters the $a_1$ propagator discussed below. 
The Lagrangian is chosen in a form analogous to Eq.~\rf{lagarp}
\[
\lag_\akstk=\frac{g_{\akstk}}{\sqrt{2}}
\left(\lag_1^\prime\cos\theta+\lag_2^\prime\sin\theta\right),
\]
with
\begin{eqnarray*}
\lag_1^\prime &=& \partial^\nu K^\dagger A^\mu K^*_{\mu\nu}
+ \mathrm{H.c.}\ ,\\
\lag_2^\prime &=& K^\dagger\partial^\mu A^\nu K^*_{\mu\nu}
+ \mathrm{H.c.}
\end{eqnarray*}
Matrix notation is now used, in which 
\begin{eqnarray*}
K&=&\left(\barr{c}
K^+\\K^0
\earr \right),\ 
K^*_\mu=\left(\barr{c}
K^{*+}_\mu\\K^{*0}_\mu
\earr \right),\\ 
A^\mu&=&\left(\barr{cc}
(a_1^0)^\mu & \sqrt{2}(a_1^+)^\mu\\
\sqrt{2}(a_1^-)^\mu& -(a_1^0)^\mu
\earr\right),
\end{eqnarray*}
and $K^*_{\mu\nu}=\partial_\mu K^*_\nu-\partial_\nu K^*_\mu$. As usual
\cite{hokim},
a particle symbol denotes the field operator which annihilates
that particle and creates its antiparticle. 
In the spirit of the SU(3) symmetry, we assume the same mixing 
angle $\theta$ as in the $\arp$ case \rf{lagarp}. 
The coupling constant $g_\akstk$ cannot be reliably extracted
from the experimental data yet because of conflicting information
about the $a_1\ra K\bar{K}\pi$ branching fractions.\footnote{See
\cite{pdg2008} and the discussion on p. 253 in \cite{aleph2005}.}. 
We will therefore use the SU(3) symmetry relation
\be
\label{Kstratio1}
g^2_\akstk=\frac{1}{4}g^2_\arp.
\ee

In order to evaluate the amplitudes of the Feynman diagrams depicted
in Figs \ref{fig:fig1} and \ref{fig:fig2}, we also need to 
specify the interaction Lagrangian among the $a_1$, $\pi$, and $\sigma$ 
fields. We write it in the form
\[
\lag_{a_1\sigma\pi}=g_1\left(\amu\cdot\partial_\mu{\fai}\right)S+g_2
\left(\amu\cdot\fai\right)\partial_\mu S\ ,
\]
where $S$ is the operator of the $\sigma$ field. The Lorentz
condition for the $a_1$ field implies that the amplitude of the decay
$a_1\ra\sigma+\pi$ is proportional to the difference 
\be
\label{ga1sigmapi}
g_{a_1\sigma\pi}=g_1-g_2.
\ee
In the $\tau$ decay diagrams, where the off-mass-shell $a_1$ resonance is
represented by its propagator \rf{a1prop}, also the terms proportional to 
\be
\label{ha1sigmapi}
h_{a_1\sigma\pi}=g_1+g_2
\ee 
contribute. There is no way of inferring the $g_{a_1\sigma\pi}$ and 
$h_{a_1\sigma\pi}$ from the hadron decay data. We will return to this 
problem later in this article. 

The interaction Lagrangian between the $\sigma$ and $\pi$ fields
is given by
\[
\lag_{\sigma\pi\pi}=g_{\sigma\pi\pi}\left(\fai\cdot\fai\right)S\ .
\]
The coupling constant $g_{\sigma\pi\pi}$ could be estimated from 
the data on the $\sigma$ mass and width \cite{e791sigma,cleosigma}.
But because this constant enters the amplitudes of the three-pion
decays of the taon multiplied by $g_{a_1\sigma\pi}$ or 
$h_{a_1\sigma\pi}$, which are both unknown, it does not have much sense. 

\subsection{\label{subsec:a1prop}Propagator of the $\mathbf{a_1}$ resonance}
We choose an analytically correct form \cite{tornqvist1987,isgurtau} of the
$a_1$ propagator featuring the running mass $M(s)$ and the
energy-dependent total width $\Gamma_{a_1}(s)$ in the denominator
\be
\label{a1prop}
-i G^{\mu\nu}_{a_1}(p) = \frac{-g^{\mu\nu}+p^\mu p^\nu/\mas}
{s-M^2_{a_1}(s)+im_{a_1}\Gamma_{a_1}(s)}\ .
\ee
The following conditions should hold
\bea
\label{cond1}
&&M^2_{a_1}(\mas)=\mas,\\
\label{cond2}
&&\frac{dM^2_{a_1}}{ds}(\mas)=0,\\
\label{cond3}
&&\Gamma_{a_1}(\mas)=\Gamma_{a_1},
\eea
where $m_{a_1}$ and $\Gamma_{a_1}$ are the nominal mass and width of
the $a_1(1260)$ resonance, respectively. 
The denominator in \rf{a1prop} is the boundary value of a function 
analytic in the complex $s$-plane ($s=p^2$) with a cut running along 
the real axis from the three-pion threshold to infinity. The running 
mass squared can therefore be obtained from a once-subtracted dispersion 
relation\footnote{In Refs. \cite{tornqvist1987,cleo2000a} an unsubtracted
dispersion relation was used.} with $\Gamma_{a_1}(s)$ as input
\be
\label{dispersion}
M^2_{a_1}(s)=M^2_{a_1}(0)-\frac{s}{\pi}{\mathrm P}\!\int_{9m_\pi^2}^\infty
\frac{m_{a_1}\Gamma_{a_1}(s^\prime)}{s^\prime(s^\prime-s)}ds^\prime\ . 
\ee
Symbol P denotes the Cauchy principal value.
We have chosen the subtraction point at $s=0$, instead of $s=\mas$ as
in \cite{isgurtau}. The advantage is that the integrand in \rf{dispersion}
contains just one singular point instead of two, what makes the evaluation
more stable and much faster. The disadvantage is that the condition
\rf{cond1} is not satisfied automatically and $M^2_{a_1}(0)$ must be 
recalculated if $m_{a_1}$ or any parameter inside $\Gamma_{a_1}(s)$
changes. 

The following decays are considered when calculating $\Gamma_{a_1}(s)$:
\bea
\label{rhopi}
a_1 &\ra& \rho+\pi\ra 3\pi \\ 
\label{kstk}
a_1 &\ra& \bar{K^*}K, K^*\bar{K} \ra K\bar{K}\pi,\\
\label{sigmapi}
a_1 &\ra& \sigma+\pi\ra 3\pi,
\eea
where the mass of the decaying $a_1$ is taken to be $\sqrt{s}$. We
neglect the interference between the amplitudes of \rf{rhopi} and
\rf{sigmapi} despite the identical final states. We argue 
that the decay $a_1\ra\rho\pi$ proceeds in the S and D orbital momentum 
states, whereas $a_1\ra\sigma\pi$ in the P state. This argument
is not entirely watertight because neither \rf{rhopi} nor \rf{sigmapi} 
satisfies the conditions for being factorized as a two-step process 
\cite{aps}.
Channel \rf{kstk} is described by four Feynman diagrams, two of them
have identical final states (\eg, $K^-K^0\pi^0$ in the case of $a_1^-$.)
We checked that the interference can be safely neglected in this case.

The Lagrangian between the vector ($K^*$)
and pseudoscalar ($K$, $\pi$) fields is chosen in a standard form
with coupling constant $g_{K^*K\pi}$. 
Moreover, the empirical widths \cite{pdg2008} of the $\rho(770)$ and 
$K^*(892)$ provide the ratio
\be
\label{Kstratio2}
\frac{g^2_{K^*K\pi}}{g^2_{\rho\pi\pi}}=0.883\pm0.035,
\ee
where the error has been enlarged to absorb the difference between the
charged and neutral $K^*(892)$. Value \rf{Kstratio2} is a little higher
than the SU(3) value of $3/4$. Relations \rf{Kstratio1} and \rf{Kstratio2}
enable us to express the product of coupling constants squared acting 
in \rf{kstk} as a multiple of 
\be
G^2=g^2_\arp g^2_{\rho\pi\pi},
\ee 
which determines the partial decay width of \rf{rhopi}.  We introduce
the ratio
\be
\label{x}
x=\frac{g^2_\akstk g^2_{K^*K\pi}}{G^2},
\ee
the value of which is given by multiplying \rf{Kstratio1} by
\rf{Kstratio2}.

As we have already mentioned, there is no way of getting the product 
$g_{a_1\sigma\pi} g_{\sigma\pi\pi}$ from the data, as the partial decay 
width of \rf{sigmapi} is unknown. We therefore proceed in another way. 
We define the parameter
\be
\label{y}
y=\frac{g_{a_1\sigma\pi}g_{\sigma\pi\pi}}{G}.
\ee
If we insert the parameters $x$ and $y$ into the formula for the $a_1$
total decay width, it becomes proportional to $G^2$. So does the derivative 
of the running mass squared \rf{dispersion}. When the condition \rf{cond2}
is applied, $G^2$ can be canceled. With known $x$, the condition \rf{cond2} 
thus becomes an equation for the unknown $y^2$. As we neglect the 
possible interference between the Feynman diagrams containing the $\rho$ 
with those containing the $\sigma$, the sign of $y$ is not essential and we 
choose $y\geq 0$. The dimension of $y$ is (energy)$^2$ because the two
Lagrangians that describe the decay $a_1\ra 3\pi$ via $\rho\pi$ have 
together three derivatives, while those via $\sigma\pi$ just one.   

An important note concerns the hadron vertexes. The effective Lagrangian 
approach takes hadrons as elementary quanta of the corresponding fields, 
ignoring thus their internal structure. As a consequence, the interaction 
strength is overestimated at higher momentum transfers. To describe the 
interaction among participating mesons more realistically, we explore 
the chromoelectric flux-tube breaking model of Kokoski and Isgur 
\cite{kokoski1987}, as it was done already in the IMR model \cite{isgurtau}.
 Each strong interaction vertex is modified by the factor 
\be
\label{fluxtube}
F(q)=\exp\left\{-\frac{q^2}{12\beta^2}\right\},
\ee
where $q$ is the three-momentum magnitude of a daughter meson in the
rest frame of the parent one (virtual masses are taken in the 
intermediate states). In the original paper \cite{kokoski1987}, the value
$\beta=0.4$~GeV/$c$ was established. We will use this value in all
our calculations, as we found that moving from it did not bring 
statistically significant improvement of the agreement of our model with data. 
A cutoff similar to \rf{fluxtube} was used in the model by CLEO Collaboration
\cite{cleo2000a}. Their parameter $R=1.2$ corresponds to $\beta=0.340$.

During the course of development of our model we tried various
versions of the $a_1$ propagators, from the most primitive one with
the constant mass and width to the most sophisticated and best physically
justified one \rf{a1prop}. The best fit to data has been provided by the 
latter.

When investigating the presence of the suspected radial recurrence
of the $a_1(1260)$, denoted as $a_1^\prime$, we supplement the $a_1$
propagator \rf{a1prop} with the term
\be
\label{a1primeprop}
-i G^{\mu\nu}_{a_1^\prime}(p) = \alpha\frac{-g^{\mu\nu}+p^\mu p^\nu/m^2_{a_1^\prime}}
{s-m^2_{a_1^\prime}+im_{a_1^\prime}\Gamma_{a_1^\prime}(s)}\ ,
\ee
where $\alpha$ is a complex parameter. We assume that the energy dependent 
total decay width $\Gamma_{a_1^\prime}(s)$ exhibits the same energy behavior 
as that of $a_1(1260)$ and write 
\[
\Gamma_{a_1^\prime}(s)=\frac{\Gamma_{a_1}(s)}
{\Gamma_{a_1}(m^2_{a_1^\prime})}\Gamma_{a_1^\prime}\ ,
\]
where $\Gamma_{a_1^\prime}=\Gamma_{a_1^\prime}(m^2_{a_1^\prime})$ is the
assumed width of the $a_1^\prime$ resonance. 

What concerns the relation to the previous models, our $a_1$ propagator
is closest to that used by the CLEO Collaboration \cite{cleo2000a}. If we 
ignored the momentum dependent terms in the numerators of \rf{a1prop} 
and \rf{a1primeprop}, we would recover their Breit-Wigner function. 

\subsection{Propagators of the $\bm{\rho}$, $\bm{K^*}$, and $\bm{\sigma}$ 
resonances}
In order to calculate the amplitudes of the taon's three-pion decays
we need also the $\rho$ and $\sigma$ propagators. They play a role also 
in decays \rf{rhopi} and \rf{sigmapi}. In addition, the 
evaluation of the decay width \rf{kstk} requires the knowledge
of the $K^*$ propagator.

We choose the propagator of both the charged and neutral rho resonances
in the form
\be
\label{rhoprop}
-i G^{\mu\nu}_{\rho}(p) = \frac{-g^{\mu\nu}+p^\mu p^\nu/\mrs}
{s-\rmrs+im_\rho\Gamma_\rho(s)}\ ,
\ee
which uses the running mass squared $\rmrs$ and the energy dependent total
width $\Gamma_\rho(s)$ from Ref. \cite{running}. The denominator of propagator
\rf{rhoprop} is an analytic function in the $s$-plane with a cut running
from $4m_\pi^2$ to infinity, as required by general principles. 
The real function $\rmrs$ is calculated from $\Gamma_\rho(s)$
using a once-subtracted dispersion relation, which guarantees that the
condition $M_\rho^2(\mrs)=\mrs$ is satisfied. The condition
\[
\frac{dM_\rho^2}{ds}(\mrs)=0
\]
is not fulfilled automatically and serves as a check that all important 
contributions to the total $\rho$-meson width $\Gamma_\rho(s)$ have 
properly been taken into account. They include, in addition to the basic 
two-pion decay channel, the $\omega\pi^0$,  $K^+K^-$, $K^0\bar K^0$, and 
$\eta\pi^+\pi^-$, which get open as the $\rho$ resonance goes above its 
nominal mass. The structure of the participating mesons is taken into
account by means of the Kokoski-Isgur form factor \rf{fluxtube}.
 
The running mass description of the $\rho$ propagator \cite{running} 
differs from other approaches that appeared in the literature 
\cite{GS,VW,melikhov04}. Gounaris and Sakurai \cite{GS} considered 
only the two-pion contribution to the total width of the $\rho^0$ 
resonance and ignored structure effects. The result is a simple
analytic formula, the main reason why their approach is so popular.
Vaughn and Wali \cite{VW} took into account the strong form factor, 
but again ignored higher decay channels. Melikhov, Nachtmann, Nikonov, 
and Paulus \cite{melikhov04} included the $K^+K^-$ and $K^0\bar K^0$ 
channels, but did not consider the strong form factors. The running 
mass formalism \cite{running} takes into account both the higher
decay channels and the structure effects.

The propagator of the $K^*(892)$ resonance is required only for the
calculation of the decay rate \rf{kstk}, which contributes to the total 
decay width of the $a_1(1260)$ resonance. It does not act in the 
three-pion decay of the $\tau$ lepton. It is chosen in a simpler form, 
with the constant mass and energy dependent decay width
\be
\label{kstprop}
-iG^{\mu\nu}_{K^*}(p) = \frac{-g^{\mu\nu}+p^\mu p^\nu/m^2_{K^*}}
{s-m^2_{K^*}+im_{K^*}\Gamma_{K^*}(s)}\ .
\ee
The decay width includes only the contribution from the $K^*\ra K+\pi$
channel and is normalized to the nominal width $\Gamma_{K^*}$ 
at $s=m^2_{K^*}$. The corresponding formula, taking into account also
the Kokoski-Isgur form factor \rf{fluxtube}, is
\[
\Gamma_{K^*}(s)=\frac{m_{K^*}^2}{s}\left[
\frac{q(s)}
{q(m^2_{K^*})}
\right]^3\frac{F(q(s))}{F(q(m^2_{K^*}))}\Gamma_{K^*},
\]
where $q(s)$ is the momentum of a daughter particle in the rest frame
of the parent $K^*$ with the mass $\sqrt{s}$. The $K^*(892)$ is a narrow
resonance and we experienced numerical instabilities when calculating
integrals containing the square of \rf{kstprop}. To get rid of problems,
we have used the procedure described in Appendix \ref{app:narrow}.

Also for the $\sigma$ propagator we use the form with fixed 
mass and energy dependent width 
\[
-i G_{\sigma}(p) = \frac{1}{s-m^2_\sigma+im_\sigma\Gamma_\sigma(s)}\ ,
\]
where $\Gamma_\sigma(s)$ includes only the contribution from the
two-pion decay channel and is equal to
\[
\Gamma_\sigma(s)=\frac{m_\sigma^2}{s}\sqrt{\frac{s-4m_\pi^2}
{m_\sigma^2-4m_\pi^2}}\frac{F(q(s))}{F(q(m^2_\sigma))}\ \Gamma_\sigma.
\]
As the current Review of Particle Physics \cite{pdg2008} is not very
specific about the $f_0(600)$ mass and width, we rely on the 
mutually compatible values obtained  by the Fermilab E791 Collaboration
\cite{e791sigma} and the CLEO Collaboration \cite{cleosigma}, who 
both analyzed the $D$ mesons decays. The results (in MeV) of E791 are
$m_\sigma=478^{+24}_{-23}\pm17$, $\Gamma_\sigma=324^{+42}_{-40}\pm21$,
whereas those of CLEO are $m_\sigma=513\pm32$, $\Gamma_\sigma=335\pm67$.
We adopt the weighted averages $m_\sigma=500$~MeV and 
$\Gamma_\sigma=329$~MeV.

\section{\label{sec:data}Experimental data}
We will compare the calculated three--pion mass distribution in the 
$\tauch$ and $\taunn$ decays with the outcome of the five experiments.

(1) The ARGUS Collaboration \cite{argus1993} used the ARGUS detector at the 
DORIS II $\die$ storage ring at DESY and studied the $\tauch$ decay. 
Their background and acceptance corrected three pion mass distribution 
is given in twenty-eight bins within the mass range \mbox{0.425--1.775 GeV}. 

(2) The OPAL Collaboration published their results on the three-pion-mass squared 
distribution in the charged-pion channel in two papers. The first of them 
\cite{opal1995} was based on the data collected with the OPAL detector at 
the CERN Large Electron-Positron Collider (LEP) during 1992 and 1993. We 
used it in our recent publication \cite{licvoj}. In this work we explore the
updated version \cite{opal1997}, in which also the data of 1994 were
included. The three-pion-mass-squared plot is corrected for background and 
efficiency and consists of twenty-three bins with much smaller statistical
errors than in \cite{opal1995}.

(3) The $\tau$-lepton decay into three pions and neutrino was also 
investigated 
by the CLEO Collaboration at the Cornell Electron Storage Ring (CESR). 
Their results on the all-charged-pions channel still exist
only in a preliminary form \cite{cleo_prelim}. We can therefore use only
the data on the $\pi^-\pi^0\pi^0$ channel \cite{cleo2000a}. The 
background-subtracted, efficiency corrected three-$\pi$ mass spectrum 
is given in 47 bins. 

(4,5) The ALEPH Collaboration at CERN LEP have published an article 
summarizing their results about the branching ratios and spectral functions
of the $\tau$ decays \cite{aleph2005}. It is based on the data collected with 
the ALEPH detector during 1991-1995 but processed by an improved method. 
We use the tables of the corrected three-pion mass squared spectra both 
in the $\tauch$ and $\taunn$ decays, which are publicly accessible at 
the website \cite{alephweb}. The all-charged pion spectrum contains 116 
bins 0.025~GeV$^2$ wide starting at 0.225 GeV$^2$. In the 
two-neutral pion case the spectrum starts at 0.2 GeV$^2$, but we discard 
the bin centered at 0.2375~GeV$^2$ with a zero value and, comparing to
its neighbors, an unrealistically small error. We are thus left again with 
116 bins. In both cases we ignore the correlation matrices among
the errors in different bins and add statistical and systematic errors 
linearly. 

\section{\label{sec:results}Calculations and results}
To be sure that our results are free of programming errors, we have written
two independent computer codes, one in \textsc{C++} (M.V.), 
another in \textsc{Fortran~95} (P.L.) and debugged them until 
they produced identical results.
 
We found that the parity-violating term in the $\tau$ decay amplitude 
influences the three-pion-mass distribution only negligibly and have 
not considered it any longer in our calculations. The decay amplitude 
$\mathcal M$ in \rf{dg4dw} then depends only on relativistic invariants
$s_{ij}=(p_i+p_j)^2$, which are symmetric against transformation
$\varphi^\prime_3\ra 2\pi-\varphi^\prime_3$. Using this symmetry when 
calculating the innermost integral in \rf{dg4dw} enables us to speed 
up the computing by a factor of two.

In the diagrams with the $\sigma$ in the intermediate states, depicted 
in Figs.~\ref{fig:fig1} and \ref{fig:fig2}, also the product 
$h_{a_1\sigma\pi}g_{\sigma\pi\pi}$ plays a role (for definitions,
see Sec.~\ref{sec:otherlags}). We introduce additional 
free parameter 
\be
\label{z}
z=\frac{h_{a_1\sigma\pi}}{g_{a_1\sigma\pi}},
\ee
which allows us to express that unknown product as a multiple of 
$g_{a_1\sigma\pi}g_{\sigma\pi\pi}$, which is determined by the
method described in Sec. \ref{subsec:a1prop}. The parameter $z$ itself 
will be obtained by minimalization of $\chi^2$ when fitting the
experimental three-pion mass distributions in the three-pion decays
of the $\tau$ lepton. 

To be closer to experimental conditions, we do not calculate the
(unnormalized) differential decay rate at certain values of the
three-pion mass $W$ or its square $Q^2\equiv W^2$, but its 
averages over the experimentally given bins in $W$ \cite{argus1993,cleo2000a}
or $Q^2$ \cite{opal1997,aleph2005}. 

When the contribution of the $a_1^\prime$ to the $a_1$ propagator is not
considered, the calculated differential decay rates, and thus also the
$\chi^2$ evaluated from them and data, depend on the following 
four parameters: (1) the nominal $a_1$ mass $m_{a_1}$, (2) the nominal 
$a_1$ width $\Gamma_{a_1}$, (3) the $\arp$ Lagrangian mixing parameter
$\sth$, and (4) the off-mass-shell coupling constant ratio $z$ 
defined by Eq.~\rf{z}. Quantity $y$ \rf{y} is not an extra parameter,
condition \rf{cond2} determines it as an implicit function of $m_{a_1}$
and $\sth$.

The ARGUS, OPAL, and CLEO experiments present the three-pion mass spectra 
in the acceptance corrected number of events. Both ALEPH spectra are
normalized to the integrated branching fractions. As the outcome of our 
model is not normalized (the coupling of the $a_1$ meson to the $W$ boson
is not fixed by meson dominance \cite{vavd}), we opt to compare just shape 
of the mass distribution and introduce five multiplicative constants. The 
values of them are obtained by minimizing the individual $\chi^2$'s 
for each experiment while keeping the common parameters ($m_{a_1}$, 
$\Gamma_{a_1}$, $\sth$, and $z$) fixed.
 
\begin{figure}
\includegraphics[width=8.6cm]{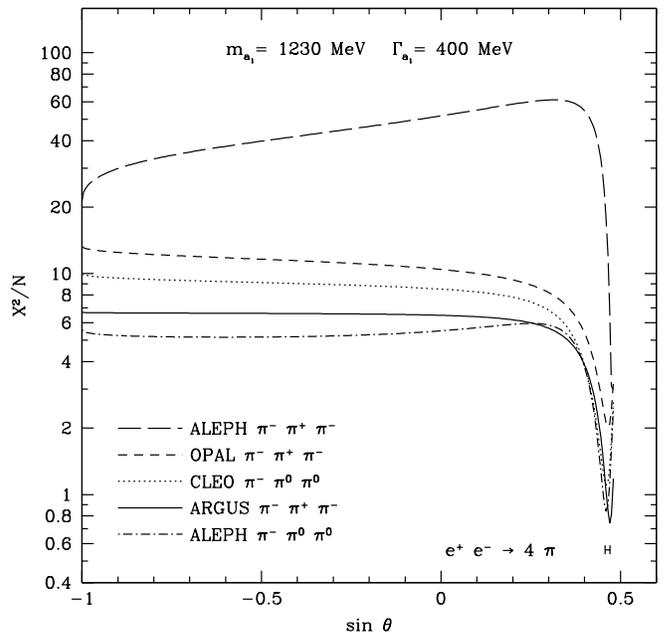}
\caption{\label{fig:fig3}Dependence of $\chi^2$ divided
by the number of experimental points on the Lagrangian mixing parameter
$\sth$ for individual data sets. The interval of $\sth$  suggested
by the electron-positron annihilation into four pions \cite{licjur,jurlic} is 
depicted as a short abscissa in the bottom right corner.}
\end{figure}

To get a quick insight into the dependence of the quality of the fit 
on the Lagrangian mixing parameter $\sth$, we first fix the basic $a_1$ 
parameters at the ``standard'' values, frequently used in theoretical 
considerations, namely, $m_{a_1}=1.23$~GeV/$c^2$ and $\Gamma_{a_1}=0.4$~GeV. 
We also set $z=0$ and calculate the ratio of the usual $\chi^2$ to the 
number of experimental points $N$ for each of the five data sets as a
function of $\sth$. The results are shown in Fig.~\ref{fig:fig3}. 
It is clear that the choice of the correct $\arp$ Lagrangian is of the 
utmost importance for obtaining a good agreement with the data. The fact 
that all five experiments point to the same narrow region in $\sth$ is 
extremely important. In addition, this region overlaps with the interval 
$(0.457,0.471)$ based on the results of the model \cite{licjur,jurlic} 
of the electron-positron annihilation into four pions built around the 
same Lagrangian \rf{lagarp}. It indicates the soundness both of the 
present model and of the $\die$ annihilation model. 

The region $\sth\gtrsim 0.5$ is not shown in Fig.~\ref{fig:fig3}
because for those values of $\sth$ it is impossible to satisfy condition 
\rf{cond2} by procedure described in Sec.~\ref{subsec:a1prop}. The 
square of parameter $y$, defined by Eq.~\rf{y}, acquires negative, 
i.e. unphysical, values, which mean the negative branching ratio of 
decay \rf{sigmapi}. 

In the next step we allow all four parameters to vary and use the CERN
computer library program \textsc{Minuit} of James and Roos \cite{minuit}
for finding their values that minimize $\chi^2$ for the three data
sets defined in Sec.~\ref{sec:data}. The results are summarized in
Table \ref{tab:results}. As always, the assessment of errors of the
\begin{table*}
\caption{\label{tab:results}%
Results of fitting various data sets. Only $a_1(1260)$
considered. Parameter $y$ is defined by Eq.~\rf{y}, parameter $z$ 
by Eq.~\rf{z}. All= ARGUS
\cite{argus1993} + OPAL \cite{opal1997} + CLEO \cite{cleo2000a} + 
both ALEPH \cite{aleph2005} data sets. For comparison, the values of 
Lagrangian mixing parameter from the $\die$ annihilation into four 
pions are also shown.}
\begin{ruledtabular}
\begin{tabular}{lcrrccccc}
Data  & Type&$\chi^2$/NDF & C.L. (\%)&$m_{a_1}$(MeV)&
$\Gamma_{a_1}$(MeV)&$\sin\theta$ &$y$ (GeV$^2$)& $z$\\
\colrule
ALEPH \cite{aleph2005}&$\tripi$&119.1/111& 28.25 & $1220\pm20$ & 
$418\pm40$ & $0.460\pm0.004$&$0.094\pm0.010$&$0.31\pm0.03$\\
ALEPH \cite{aleph2005}&$\pitwo$& 51.5/111&100.00 & $1256\pm10$ & 
$443\pm15$ & $0.466\pm0.004$&$0.111\pm0.022$&$0.12\pm0.16$\\
All                   & Mixed  &357.7/321&  7.74 & $1232\pm25$ & 
$431\pm25$ & $0.463\pm0.005$&$0.099\pm0.009$&$0.30\pm0.05$\\
\colrule
 & \multicolumn{5}{l}{$e^+e^-\ra\pi^+\pi^-\pi^+\pi^-$ \cite{licjur}} &
 $0.460\pm0.003$ \\
 & \multicolumn{5}{l}
{$e^+e^-\ra\pi^+\pi^-\pi^+\pi^-\  \&\  \pi^+\pi^-\pi^0\pi^0$
 \cite{jurlic}} & $0.466\pm0.005$ \\
\end{tabular}
\end{ruledtabular} 
\end{table*}
parameters is a difficult task. We combined the errors provided
by \textsc{Minuit}, which  reflect the errors of experimental data,
with our estimates of the errors induced by the uncertainties of the input
parameters (the $\sigma$ mass, various coupling constants). The agreement
of our model with the ALEPH $\pitwo$ data is perfect (confidence level
100\%), the agreement with two other data sets is satisfactory by measures
usually accepted in the high energy physics ($\chi^2$/NDF $\approx 1$). 
The values of the $a_1$ mass obtained from the three data sets are
mutually compatible, as well as those of the $a_1$ width. What is
especially remarkable are the values of the Lagrangian mixing parameter 
$\sth$. Not only their small errors ($\approx 1\%$) and mutual consistence,
but also a perfect agreement with the values obtained from the analyzes
of the $\die$ annihilation into four pions. Parameter $z$ characterizes
the part of the $a_1\sigma\pi$ Lagrangian that acts only for virtual
$a_1$ and cannot be compared with anything yet (the $\die$ annihilation
model \cite{licjur,jurlic} did not consider the $\sigma\pi$ intermediate
states).

Now we add the $a_1^\prime$ contribution \rf{a1primeprop} to the $a_1$
propagator \rf{a1prop}. Not to increase the number of free parameters too
much, we fix the mass and width of the $a_1^\prime$ at the PDG 2008 values
1647~MeV and 254~MeV, respectively. The same approach was used by the
CLEO Collaboration \cite{cleo2000a}, just their values were a little
different (1700~MeV and 300~MeV). The number of the free parameters thus
increases by two [the real and imaginary parts of $\alpha$, see
\rf{a1primeprop}]. The results of the $\chi^2$ minimalization procedure are
shown in Table~\ref{tab:a1primeresults} for all three data sets.
\begin{table*}
\caption{\label{tab:a1primeresults}%
Results of fitting various data sets. Both $a_1(1260)$ and $a_1(1640)$ 
are considered. For definition of $\alpha$, see Eq.~\rf{a1primeprop}.}
\begin{ruledtabular}
\begin{tabular}{lcrcccccccc}
Data & Type &$\chi^2/$NDF&C.L.& $m_{a_1}$(MeV)& $\Gamma_{a_1}$(MeV)
&$\sin\theta$&$y$~(GeV$^2)$ & $z$ & Re~$\alpha$& Im~$\alpha$\\
\colrule
ALEPH \cite{aleph2005}&$\tripi$& 30.7/109&100\% &$1218\pm19$&$418\pm30$ &
0.457(4)&$0.106\pm0.019$&$0.34\pm0.03$&$-0.30\pm0.10$&$0.31\pm0.06$\\
ALEPH \cite{aleph2005}&$\pitwo$& 12.3/109&100\% &$1255\pm18$&$455\pm15$ &
0.457(6)&$0.148\pm0.025$&$0.36\pm0.14$&$-0.34\pm0.13$&$0.29\pm0.10$\\
All                   & Mixed  &219.5/318&100\% &$1233\pm18$&$431\pm20$ & 
0.459(4)&$0.114\pm0.014$&$0.34\pm0.05$&$-0.31\pm0.10$&$0.32\pm0.09$\\
\end{tabular}
\end{ruledtabular}
\end{table*}
The comparison of Tables \ref{tab:results} and \ref{tab:a1primeresults}
shows that the addition of the $a_1^\prime$ resonance to the $a_1$ propagator
greatly improves the agreement of the model with data in all cases.
For the all-charged-pions ALEPH data 
\cite{aleph2005}, the $\chi^2$ drops from 119.1 to 30.7 and the confidence 
level rockets from 28.25\% to 100\%. The improvement of the confidence 
level is even more substantial for the third data set, where the total
$\chi^2$ is a sum of the individual $\chi^2$'s for the ARGUS, OPAL,
CLEO, ALEPH $\tripi$, and ALEPH $\pitwo$ data. The mass and width of
the $a_1$ as well as other two free parameters ($\sth$ and $z$)
are very stable against the inclusion of $a_1^\prime$. Their new values
(Table \ref{tab:a1primeresults}) differ only very little from the
corresponding old ones (Table \ref{tab:results}). Also the values 
obtained from different data sets are mutually compatible. This is
true also for two new parameters Re~$\alpha$ and Im~$\alpha$. 

\begin{figure}
\includegraphics[width=8.6cm]{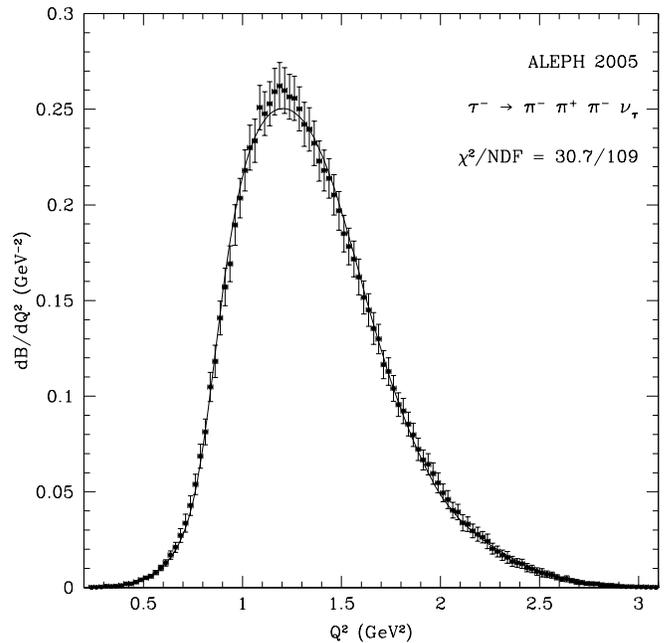}
\caption{\label{fig:aleph_3pi}Three-pion-mass-squared distribution 
calculated assuming both $a_1$ and $a_1^\prime$ contributions and
compared to the ALEPH \cite{aleph2005} $\tauch$ data. The model 
parameters taken from Table~\ref{tab:a1primeresults}.}
\end{figure}

\begin{figure}
\includegraphics[width=8.6cm]{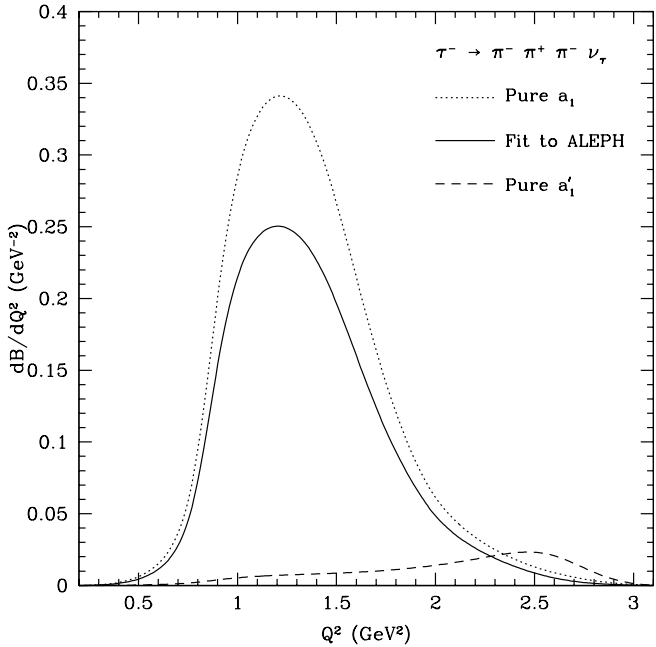}
\caption{\label{fig:fig5}Investigating the role of the $a_1^\prime$
resonance in fitting the ALEPH $\tripi$ data. Full curve: the complete 
calculation shown in Fig.~\ref{fig:aleph_3pi}; Dotted curve: parameters 
unchanged, but only the $a_1(1260)$ term \rf{a1prop} in the $a_1$ propagator; 
Dashed curve: parameters unchanged, but only the $a_1^\prime$ term 
\rf{a1primeprop} in the $a_1$ propagator. Note the change of scale 
against Fig.~\ref{fig:aleph_3pi}.}
\end{figure}

\begin{figure}
\includegraphics[width=8.6cm]{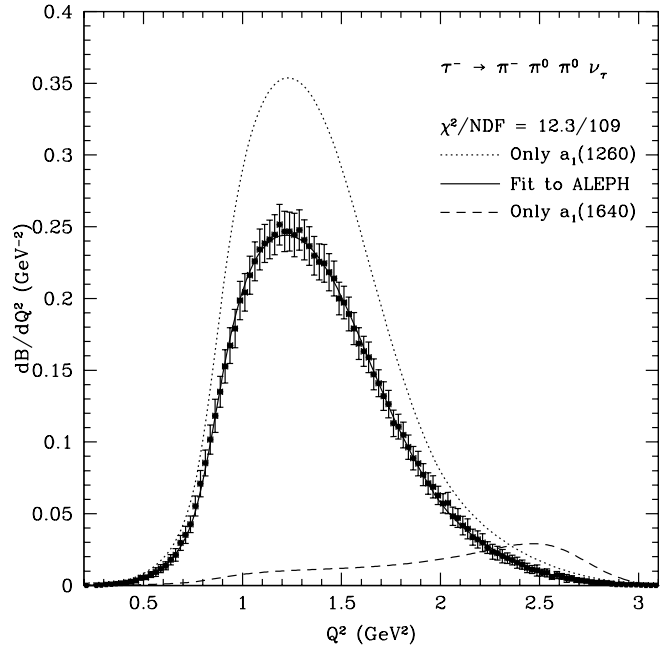}
\caption{\label{fig:fig6}Investigating the role of the 
$a_1^\prime$
resonance in fitting the ALEPH $\pitwo$ data. Full curve (buried
in data): the complete calculation; Dotted curve: parameters unchanged, but
only the $a_1(1260)$ term \rf{a1prop} in the $a_1$ propagator; Dashed curve:  
parameters unchanged, but only the $a_1(1640)$ term \rf{a1primeprop} in 
the $a_1$ propagator.}
\end{figure}

The calculated three-pion-mass distribution is compared to the $\tripi$
data of ALEPH Collaboration \cite{aleph2005} in Fig.~\ref{fig:aleph_3pi}. 
We have just learned that the $a_1(1640)$ resonance greatly improves the 
agreement with data. It is therefore a little surprising that there is 
no bump or shoulder corresponding to this resonance visible in 
Fig.~\ref{fig:aleph_3pi}. 
To investigate this conundrum we calculate the model distribution in three
cases: (1) both $a_1(1260)$ and $a_1(1640)$ terms in the $a_1$
propagator (this is the curve presented already in Fig.~\ref{fig:aleph_3pi}); 
(2) only the $a_1(1260)$ term \ref{a1prop} in $a_1$ the propagator; (3)  
only the $a_1^\prime$ term \rf{a1primeprop}. The model parameters in all
three cases are identical. They are taken from the ALEPH $\tripi$ row 
of Table \ref{tab:a1primeresults}. 
The findings, see Fig.~\ref{fig:fig5}, show that the underlying 
mechanism leading to the agreement with data is somewhat surprising.
The final distribution is a result of the destructive interference
between the dominant amplitude containing the $a_1(1260)$ 
propagator \rf{a1prop} and the amplitude containing the $a_1^\prime$
propagator \rf{a1primeprop}.
Similar analysis performed for the $\taunn$ decay, see 
Fig.~\ref{fig:fig6}, leads to the same conclusion.

\section{\label{sec:summary}Summary and conclusions}
\begin{figure}
\includegraphics[width=8.6cm]{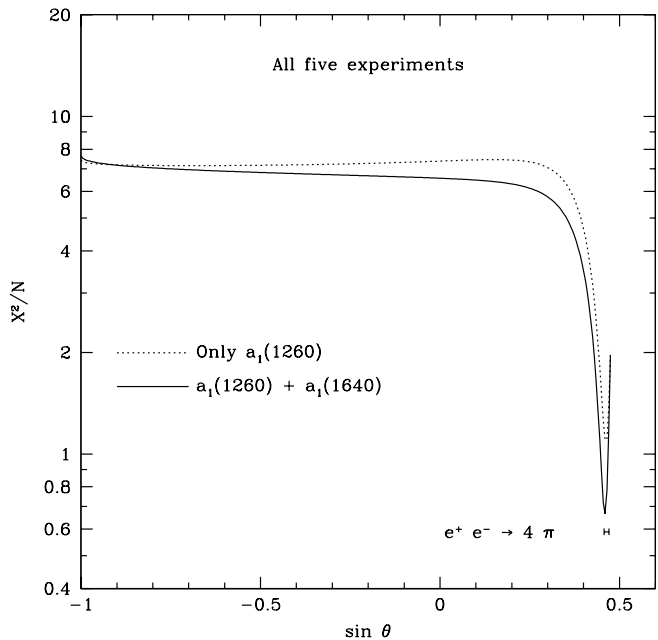}
\caption{\label{fig:fig7}$\sin\theta$ dependence of the sum
of $\chi^2$s from all five experiments divided by the total number 
of experimental points. The other parameters are kept at the optimal
values from the ``All'' row of the corresponding tables. Dotted curve: only 
$a_1(1260)$ (Table~\ref{tab:results}) included in the $a_1$ propagator; 
Full curve: both $a_1(1260)$ and $a_1(1640)$ included 
(Table~\ref{tab:a1primeresults}). The range of $\sth$ from the 
electron-positron annihilation into four pions \cite{licjur,jurlic} 
is shown as a short abscissa.}
\end{figure}

The main message of this study is that the form of the $\arp$ Lagrangian
is the decisive factor for achieving a good model description of the
three-pion decays of the tau lepton. This is again illustrated in
Fig.~\ref{fig:fig7} where the total $\chi^2$, which is calculated 
as a sum of the individual $\chi^2$'s from the five experiments, is divided
by the total number of experimental points and plotted as a function of
the Lagrangian mixing parameter $\sth$. Two different cases are considered:
(1) only $a_1(1260)$ included in the $a_1$ propagator, (2) both $a_1(1260)$ 
and $a_1(1640)$ included. In contrast to Fig.~\ref{fig:fig3},
the other parameters are fixed at their optimal values taken from the 
appropriate tables (Tabs. \ref{tab:results} and \ref{tab:a1primeresults}).
Even if the curve (2) is shifted a little toward smaller values of $\sth$,
the minima of both curves fall to the interval found in the model
of the $\die$ annihilation into four pions \cite{licjur,jurlic}.

Our further finding, even not documented in this work in detail, concerns the
form of the $a_1$ propagator. We have found that the running mass form
\rf{a1prop}, suggested and already used in several papers 
\cite{tornqvist1987,isgurtau,cleo2000a}, provides a better fit to the taon 
three-pion decay data than simpler forms with a constant $a_1$ mass and
a constant or energy dependent $a_1$ total decay width. 

The $a_1$ running mass squared is given by the dispersion relation. In this 
work we have chosen a once-subtracted version \rf{dispersion}. As the
input for the dispersion relation, the energy dependent total decay width 
of the $a_1$ for all $s$ above the three-pion threshold is required 
\rf{dispersion}. We approximated it as a sum of the decay widths to the 
three pion final states (via the $\pi\rho$ and $\pi\sigma$ intermediate 
states) and the $K\bar{K}\pi$ final states (via $K^*(892)\bar{K}$ + c.c.). 
A typical behavior of the energy dependent total width
\begin{figure}
\includegraphics[width=8.6cm]{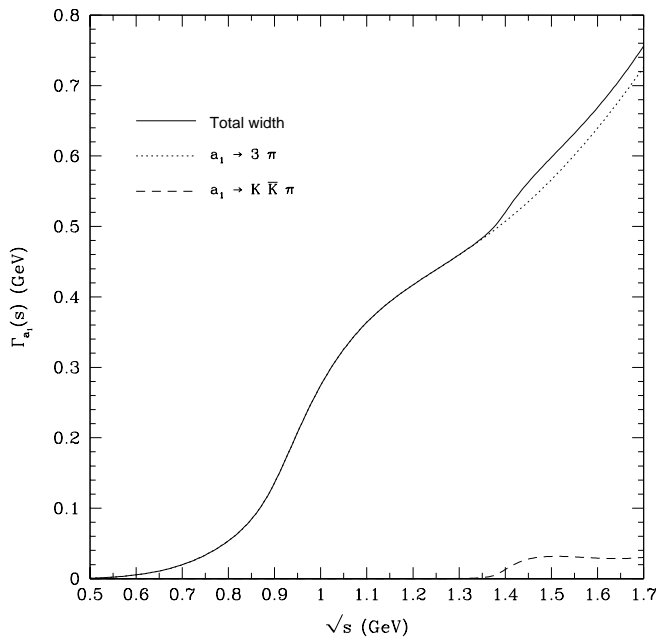}
\caption{\label{fig:fig8}Energy dependent width of the 
$a_1(1260)$ as a function of $\sqrt{s}$. Parameters $m_{a_1}$, 
$\Gamma_{a_1}$, and $\sth$ taken from Table~\ref{tab:a1primeresults}, 
row ``All''.}
\end{figure}
is shown in Fig.~\ref{fig:fig8}. The hump centered around
$\sqrt{s}\approx 1$~GeV develops as the mass of the two-pion subsystem
falls predominantly first on the ascending and then on the descending side 
of the rho propagator. We ignored 
the channels $\rho(1450)\pi$, $f_0(1370)\pi$, and $f_2(1270)\pi$, which
have been seen in the $a_1$ decays \cite{pdg2008} and which open at higher
$s$.\footnote{The inclusion of them would bring additional free parameters,
what we wanted to avoid.} This is probably the reason why the running mass 
behaves wildly, see Fig.~\ref{fig:fig9}, and does not 
have a nice plateau around the nominal mass, as it did in the case 
of the $\rho(770)$ \cite{running}. 
\begin{figure}
\includegraphics[width=8.6cm]{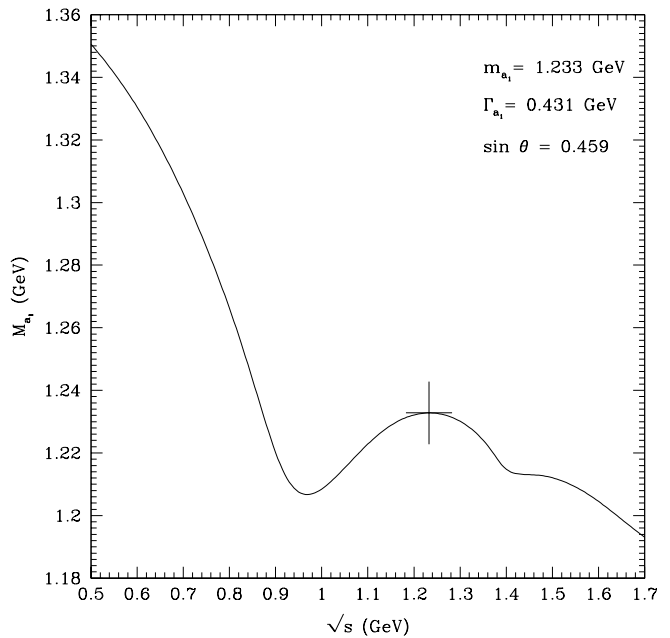}
\caption{\label{fig:fig9}Running mass of the $a_1(1260)$ as a 
function of $\sqrt{s}$ for the same parameters as 
Fig.~\ref{fig:fig8}. The cross marks the point
in which condition \rf{cond1} is satisfied. }
\end{figure}

Another important ingredient of our model are the $\pi\sigma$ intermediate
states. On one side, they enter the calculation of the total decay width 
of the $a_1$ resonance, which is necessary for constructing the running 
mass propagator \rf{a1prop}. On the other side, they contribute to the
decay rates of the three-pion decays of the tau lepton, 
Figs. \ref{fig:fig1} and \ref{fig:fig2}.
 
\begin{figure}
\includegraphics[width=8.6cm]{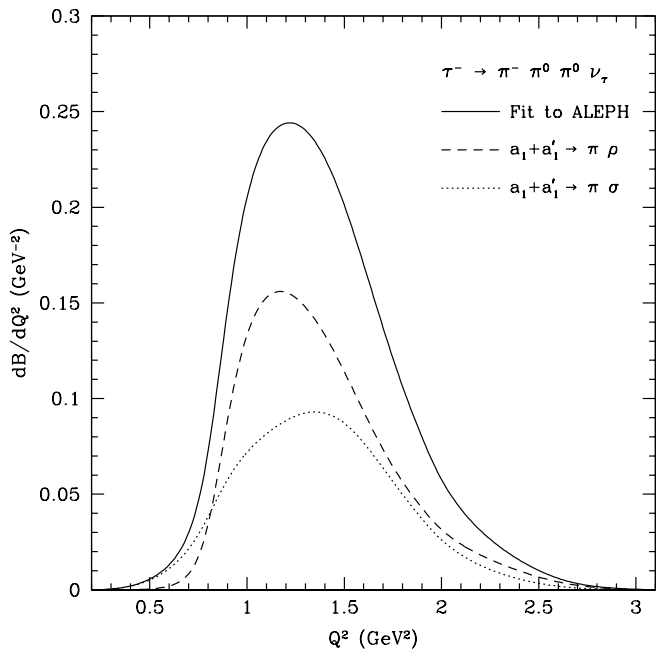}
\caption{\label{fig:fig10}Decomposition of the 
three-pion mass squared distribution into the contributions from the
$\pi\rho$ and $\pi\sigma$ intermediate states. The parameters were taken 
from the ALEPH $\pitwo$ row of Table \ref{tab:a1primeresults}.}
\end{figure}

To investigate the role of the $\pi\sigma$ intermediate states in the 
evaluation of the $\taunn$ decay width, 
we split the distribution depicted in Fig.~\ref{fig:fig6} by
the full curve into its $\pi\rho$ and $\pi\sigma$ components. The
result is shown in Fig.~\ref{fig:fig10}.
It is obvious that the $\pi\sigma$ intermediate states play a unique
role in describing the behavior of the differential decay width
at small three-pion masses. What is a little suspicious, is the large
magnitude at the intermediate masses. To see whether it is reasonable or 
not, we integrate the distributions to get the branching ratio
\[
\mathcal{B}=\frac%
{\Gamma(\tau^-\ra\nu_\tau\pi^-\sigma\ra\nu_\tau\pi^-\pi^0\pi^0)}
{\Gamma(\tau^-\ra\nu_\tau\pi^-\pi^0\pi^0)}
\]
with the result $\mathcal{B}\approx 41\%$. This number is more than
twice higher than the experimental value of $(16.18\pm3.85\pm1.28)\%$
obtained by the CLEO Collaboration (Ref. \cite{cleo2000a}, Table III).

The source of this obvious deficiency of our model is the following: The 
important parameter $y$, which regulates the rate of the 
$a_1\ra\pi\sigma\ra 3\pi$ transition, was obtained from the condition that 
the derivative of the running mass at the nominal-mass point should vanish 
\rf{cond2}. But the absence of higher decay channels, which influence the values of the 
running mass at all $s$, may modify the resulting value of $y$ 
significantly. The larger than correct $y$ may mimic the missing channels.

A new feature of our work, which, to our knowledge, has not appeared
in the literature yet, is that we fit the data from several 
experiments simultaneously. We intend to continue in this approach
and include not only the data concerning the three-pion decays of the
tau lepton, but also the experimental results from other weak, 
electromagnetic, and perhaps also strong interaction processes. The natural 
candidate is the electron-positron annihilation into four pions, for which 
a model based on the same Lagrangian as here has already been built.
The value of the Lagrangian mixing parameter we have obtained here
perfectly agrees with values obtained from
the $\die$ annihilation into four charged pions \cite{licjur} and from the
combined fit to both annihilation channels \cite{jurlic}.

To summarize:\\
(1) We have shown that the right form of the $\arp$ Lagrangian is extremely
important for obtaining a good agreement with data. We have obtained an
unprecedented confidence level of 100\% for all three sets of data we
considered. The optimal value of the Lagrangian mixing parameter $\sth$
perfectly agrees with the value obtained from the $\die$ annihilation
into four pions.\\ 
(2) Our confirmation of the existence of the $a_1(1640)$ resonance with
the mass and width compatible with the PDG \cite{pdg2008} values is
based on the increase of the confidence level from 7.7\% to 100\%
after the $a_1(1640)$ has been included.\\
(3) We have explained why the $a_1(1640)$ resonance, which is important for
getting a good agreement with data, is not visible in the three-pion-mass
spectrum as a bump or shoulder.\\
(4) From the common fit to the data from five experiments we have obtained
the following results:

Mass of the $a_1(1260)$ $m_{a_1}=(1233\pm18)$~MeV;
 
Width of the $a_1(1260)$ $\Gamma_{a_1}=(431\pm20)$~MeV;

Lagrangian mixing parameter $\sth=0.459\pm0.004$.


\begin{acknowledgements}
One of us (P. L.) is indebted to J. Kapusta for discussions many years ago 
that triggered this investigation. This work was supported by the Czech 
Ministry of Education, Youth and Sports under contracts LC07050 and 
MSM6840770029.
\end{acknowledgements}

\appendix  
\section{\label{app:newformula}Differential decay rate formula}
We use the following formula for the differential decay rate in the
invariant three-particle mass $W=\sqrt{(p_2+p_3+p_4)^2}$ in a four-body 
decay $a\ra1+2+3+4$:  
\bea
\label{dg4dw}
\frac{d\Gamma}{dW}&=& \frac{|\bm{p}_1|}{16(2\pi)^6
m_a^2}\int_{m_3+m_4}^{W-m_2}\ dm_{34}|\bm{p}^*_2|\ |\bm{p}_3^\prime| \nl
&\times&
\int_{-1}^1 d\cos\theta_2^*\int_{-1}^1d\cos\theta_3^\prime
\int_0^{2\pi}d\varphi_3^\prime\left|{\cal M}\right|^2.
\eea
The asterisk denotes the (2,3,4) rest frame, the prime the (3,4) rest frame. 
$m_{34}$ is the mass of the system consisting of particles 3 and 4, 
$E_{34}^*=E_3^*+E_4^*$ and $\mathbf{P}^*_{34}=\mathbf{p}_3^*+\mathbf{p}_4^*
=-{\mathbf p}_2^*$ are its energy and momentum, respectively, 
in the (2,3,4) rest frame.  In the rest frame of the parent
particle $a$ the momentum of particle 1 points along the negative $z$-axis.
In the (2,3,4) rest frame, the momentum of particle 2 lies in the
(xz) plane. 
\section{\label{app:narrow}Integrating over a narrow peak}
Let us assume that we need to evaluate an integral over an interval that
includes a narrow resonance peak
\be
\label{q}
Q=\int_{s_1}^{s_2}\frac{f(s)}{(s-M^2(s))^2+m^2\Gamma^2(s)}\ ds\ ,
\ee
where $f(s)$ is a slowly varying function.
Further, let the two functions in the denominator satisfy conditions
$M^2(m^2)=m^2$ and $\Gamma(m^2)=\gamma$. If $\gamma\ll m$ then the 
integrand is rapidly varying function of $s$ and a numerical
quadrature of very high order is required to get reliable results. 
After introducing a new variable $\xi$ by substitution
$s=m^2+m\gamma\tan\left(c\ \xi+d\right)$,
where $c=(a_2-a_1)/2$, $d=(a_1+a_2)/2$,
$a_1=\arctan\{(s_1-m^2)/(m\gamma)\}$, and
$a_2=\arctan\{(s_2-m^2)/(m\gamma)\}$,
the integral \rf{q} becomes
\[
Q=\frac{a_2-a_1}{2m\gamma}\int_{-1}^1\frac{(s-m^2)^2+m^2\gamma^2}
{(s-M^2(s))^2+m^2\Gamma^2(s)}f(s)\ d\xi\ ,
\]
which can be safely evaluated using, \eg, the Gauss-Legendre quadrature.
We apply this method for calculating the integrals containing the square
of the $K^*(892)$ propagator \rf{kstprop}. In that case $M(s)\equiv
m=m_{K^*}$.

\end{document}